\input harvmac

\nref\BerensteinJQ{
D.~Berenstein, J.~M.~Maldacena and H.~Nastase,
``Strings in flat space and pp waves from N = 4 super Yang Mills,''
JHEP {\bf 0204}, 013 (2002)
[arXiv:hep-th/0202021].
}

\nref\SpradlinAR{
M.~Spradlin and A.~Volovich,
``Superstring interactions in a pp-wave background,''
Phys.\ Rev.\ D {\bf 66}, 086004 (2002)
[arXiv:hep-th/0204146].
}

\nref\KristjansenBB{
C.~Kristjansen, J.~Plefka, G.~W.~Semenoff and M.~Staudacher,
``A new double-scaling limit of N = 4 super Yang-Mills theory and PP-wave
strings,''
Nucl.\ Phys.\ B {\bf 643}, 3 (2002)
[arXiv:hep-th/0205033].
}

\nref\GrossSU{
D.~J.~Gross, A.~Mikhailov and R.~Roiban,
``Operators with large R charge in N = 4 Yang-Mills theory,''
Annals Phys.\  {\bf 301}, 31 (2002)
[arXiv:hep-th/0205066].
}

\nref\ConstableHW{
N.~R.~Constable, D.~Z.~Freedman, M.~Headrick, S.~Minwalla, L.~Motl,
A.~Postnikov and W.~Skiba,
``PP-wave string interactions from perturbative Yang-Mills theory,''
JHEP {\bf 0207}, 017 (2002)
[arXiv:hep-th/0205089].
}

\nref\KiemXN{
Y.~Kiem, Y.~Kim, S.~Lee and J.~Park,
``pp-wave / Yang-Mills correspondence: An explicit check,''
Nucl.\ Phys.\ B {\bf 642}, 389 (2002)
[arXiv:hep-th/0205279].
}

\nref\ChuPD{
C.~S.~Chu, V.~V.~Khoze and G.~Travaglini,
``Three-point functions in N = 4 Yang-Mills theory and pp-waves,''
JHEP {\bf 0206}, 011 (2002)
[arXiv:hep-th/0206005].
}

\nref\VerlindeIG{
H.~Verlinde,
``Bits, matrices and 1/N,''
arXiv:hep-th/0206059.
}

\nref\SpradlinRV{
M.~Spradlin and A.~Volovich,
``Superstring interactions in a pp-wave background II,''
JHEP {\bf 0301}, 036 (2003)
[arXiv:hep-th/0206073].
}

\nref\KlebanovMP{
I.~R.~Klebanov, M.~Spradlin and A.~Volovich,
``New effects in gauge theory from pp-wave superstrings,''
Phys.\ Lett.\ B {\bf 548}, 111 (2002)
[arXiv:hep-th/0206221].
}

\nref\GursoyYY{
U.~Gursoy,
``Vector operators in the BMN correspondence,''
arXiv:hep-th/0208041.
}

\nref\BeisertBB{
N.~Beisert, C.~Kristjansen, J.~Plefka, G.~W.~Semenoff and M.~Staudacher,
``BMN correlators and operator mixing in N = 4 super Yang-Mills theory,''
Nucl.\ Phys.\ B {\bf 650}, 125 (2003)
[arXiv:hep-th/0208178].
}

\nref\SchwarzBC{
J.~H.~Schwarz,
``Comments on superstring interactions in a plane-wave background,''
JHEP {\bf 0209}, 058 (2002)
[arXiv:hep-th/0208179].
}

\nref\PankiewiczGS{
A.~Pankiewicz,
``More comments on superstring interactions in the pp-wave background,''
JHEP {\bf 0209}, 056 (2002)
[arXiv:hep-th/0208209].
}

\nref\GrossMH{
D.~J.~Gross, A.~Mikhailov and R.~Roiban,
``A calculation of the plane wave string Hamiltonian from N = 4
super-Yang-Mills theory,''
arXiv:hep-th/0208231.
}

\nref\ZhouMI{
J.~G.~Zhou,
``pp-wave string interactions from string bit model,''
Phys.\ Rev.\ D {\bf 67}, 026010 (2003)
[arXiv:hep-th/0208232].
}

\nref\ConstableVQ{
N.~R.~Constable, D.~Z.~Freedman, M.~Headrick and S.~Minwalla,
``Operator mixing and the BMN correspondence,''
JHEP {\bf 0210}, 068 (2002)
[arXiv:hep-th/0209002].
}

\nref\VamanKA{
D.~Vaman and H.~Verlinde,
``Bit strings from N = 4 gauge theory,''
arXiv:hep-th/0209215.
}

\nref\EynardDF{
B.~Eynard and C.~Kristjansen,
``BMN correlators by loop equations,''
JHEP {\bf 0210}, 027 (2002)
[arXiv:hep-th/0209244].
}

\nref\JanikBD{
R.~A.~Janik,
``BMN operators and string field theory,''
Phys.\ Lett.\ B {\bf 549}, 237 (2002)
[arXiv:hep-th/0209263].
}

\nref\PearsonZS{
J.~Pearson, M.~Spradlin, D.~Vaman, H.~Verlinde and A.~Volovich,
``Tracing the string: BMN correspondence at finite $J^2/N$,''
arXiv:hep-th/0210102.
}

\nref\GomisWI{
J.~Gomis, S.~Moriyama and J.~Park,
``SYM description of SFT Hamiltonian in a pp-wave background,''
arXiv:hep-th/0210153.
}

\nref\PankiewiczTG{
A.~Pankiewicz and B.~Stefa\'nski, jr.,
``pp-wave light-cone superstring field theory,''
Nucl.\ Phys.\ B {\bf 657}, 79 (2003)
[arXiv:hep-th/0210246].
}

\nref\BakKU{
D.~Bak and M.~M.~Sheikh-Jabbari,
``Strong evidence in favor of the existence of S-matrix for strings in
plane waves,''
JHEP {\bf 0302}, 019 (2003)
[arXiv:hep-th/0211073].
}

\nref\HeZU{
Y.~H.~He, J.~H.~Schwarz, M.~Spradlin and A.~Volovich,
``Explicit formulas for Neumann coefficients in the plane-wave geometry,''
arXiv:hep-th/0211198.
}

\nref\KiemPB{
Y.~Kiem, Y.~Kim, J.~Park and C.~Ryou,
``Chiral primary cubic interactions from pp-wave supergravity,''
JHEP {\bf 0301}, 026 (2003)
[arXiv:hep-th/0211217].
}

\nref\RoibanXR{
R.~Roiban, M.~Spradlin and A.~Volovich,
``On light-cone SFT contact terms in a plane wave,''
arXiv:hep-th/0211220.
}

\nref\GursoyFJ{
U.~Gursoy,
``Predictions for pp-wave string amplitudes from perturbative SYM,''
arXiv:hep-th/0212118.
}

\nref\MinahanVE{
J.~A.~Minahan and K.~Zarembo,
``The Bethe-ansatz for N = 4 super Yang-Mills,''
JHEP {\bf 0303}, 013 (2003)
[arXiv:hep-th/0212208].
}

\nref\BeisertFF{
N.~Beisert, C.~Kristjansen, J.~Plefka and M.~Staudacher,
``BMN gauge theory as a quantum mechanical system,''
Phys.\ Lett.\ B {\bf 558}, 229 (2003)
[arXiv:hep-th/0212269].
}

\nref\GomisKJ{
J.~Gomis, S.~Moriyama and J.~Park,
``SYM description of pp-wave string interactions: Singlet sector and
arbitrary impurities,''
arXiv:hep-th/0301250.
}

\nref\BobkovVG{
K.~Bobkov,
``Graviton-scalar interaction in the pp-wave background,''
arXiv:hep-th/0303007.
}

\nref\BeisertTQ{
N.~Beisert, C.~Kristjansen and M.~Staudacher,
``The dilatation operator of N = 4 super Yang-Mills theory,''
arXiv:hep-th/0303060.
}

\font\cmss=cmss10
\font\cmsss=cmss10 at 7pt
\def\IZ{\relax\ifmmode\mathchoice
{\hbox{\cmss Z\kern-.4em Z}}{\hbox{\cmss Z\kern-.4em Z}}
{\lower.9pt\hbox{\cmsss Z\kern-.4em Z}}
{\lower1.2pt\hbox{\cmsss Z\kern-.4em Z}}\else{\cmss Z\kern-.4em
Z}\fi}

\Title{\vbox{\baselineskip12pt
	\hbox{hep-th/0303220}
	\hbox{NSF-KITP-03-22}
	\hbox{PUPT-2079}
}}{Note on Plane Wave Quantum Mechanics}

\centerline{Marcus Spradlin${}^{1}$ and Anastasia Volovich${}^{2}$}

\bigskip
\centerline{${}^{1}$~Department of Physics}
\centerline{Princeton University}
\centerline{Princeton, NJ 08544}
\centerline{}
\centerline{${}^{2}$~Kavli Institute for Theoretical Physics}
\centerline{University of California}
\centerline{Santa Barbara, CA 93106}

\bigskip
\bigskip
\bigskip

\centerline{\bf Abstract}

\bigskip

We study the quantum mechanics of BMN operators with two scalar
impurities and arbitrarily many traces, at one loop and all genus.
We prove an operator identity which partially
elucidates the structure of this quantum mechanics, provides
some support 
for a conjectured formula for the free all genus two-point
functions, and
demonstrates
that a single ${\cal{O}}(g_2^2)$ contact term arises in the
Hamiltonian as a result of transforming from the natural
gauge theory basis to the
string basis.
We propose to identify the S-matrix of this quantum mechanics
with the S-matrix of string theory in the plane-wave background.

\smallskip

\Date{}

\listtoc
\writetoc

\newsec{Introduction}

Recently there has been much interest in a particular
limit of the AdS/CFT correspondence in which the AdS${}_5 \times
S{}^5$ background degenerates into a plane-wave and the free
string theory becomes exactly solvable \refs{\BerensteinJQ-\BeisertTQ}.
To reach this limit one
focuses in the large $N$ limit of
${\cal{N}}=4$ SU($N$) super Yang-Mills theory on those operators
with large R-charge
$J$ but finite $\Delta - J$,
where $\Delta$ is the scaling dimension \BerensteinJQ.  When
$J$ is taken to be of order $\sqrt{N}$, many quantities of
physical interest are believed to be effectively
perturbative in the parameters \refs{\BerensteinJQ,\KristjansenBB,\ConstableHW}
\eqn\aaa{
\lambda' \equiv {g_{\rm YM}^2 N \over J^2}, \qquad
g_2 \equiv {J^2 \over N},
}
despite the fact that the 't Hooft coupling $\lambda = g_{\rm YM}^2 N$ is
going to infinity in this limit.

Recent efforts have successfully matched certain matrix elements
of $\Delta - J$ in the gauge theory \refs{\BeisertBB, \GrossMH,
\ConstableVQ}, 
after a suitable basis transformation
(to be discussed extensively below), with matrix elements of
the light-cone string field theory Hamiltonian \refs{\SpradlinAR, \SpradlinRV}, 
to first
(or second) order in
the string coupling and first order in $\lambda'$ \refs{\PearsonZS, 
\GomisWI, \RoibanXR}.
It has been emphasized in a number of papers that many interesting
aspects of the gauge theory can be studied via a simple quantum
mechanical model \refs{\VerlindeIG,\GrossMH,\VamanKA,\BeisertFF, \BeisertTQ}:
there is a space of states (gauge theory operators),
an inner product (the free gauge theory two-point function),
and a Hamiltonian (given by $H=\Delta - J$ in radial quantization).

In this note we study the simplest non-trivial version of this
quantum mechanics \BeisertFF:
that of BMN operators with two bosonic impurities that are
orthogonal scalar fields, but have arbitrarily many traces.
Everything in the bulk of the paper
applies at one loop (i.e., ${\cal{O}}(\lambda')$),
although we comment on higher loops in the final section.
It is intended that this note can be read abstractly as a study
of a particular quantum mechanics
by those who are not necessarily familiar with the gauge
theory, string field theory, or the string bit model.
However, it is hoped that the results of this paper will serve to
tie together these threads a little bit.

Of particular importance in the quantum mechanics is the
splitting-joining operator
$\Sigma$  (so-called because it can increase or decrease the
number of traces by one) \VerlindeIG.
It has been conjectured that the free two-point functions of
BMN operators $\langle 1 | 2 \rangle_{g_2}$ at finite $g_2$
are related
to  those at $g_2 = 0$ by the
identity \VamanKA\
\eqn\conjone{
\langle 1 | S^{-1}| 2 \rangle_{g_2}
= \langle 1 |2\rangle_{g_2 = 0}, \qquad
S = e^{g_2 \Sigma}.
}
This formula has been confirmed by several calculations and
refuted by none so far.  If it turns out to be incorrect, then
the results of this paper remain true, albeit difficult to interpret.

In light of \conjone, we identify $S^{-1/2}$ as the transformation
between the `gauge theory' basis and the `string' basis \PearsonZS.
That is, if $|k \rangle$ is a state which corresponds to a
$k$-trace operator, then we identify $S^{-1/2} |k\rangle$ as a state
with precisely $k$ strings.
Moreover if $|k\rangle$ is the gauge theory operator which
corresponds via the BMN dictionary at $g_2=0$ to the string
state $|\widetilde{k}\rangle$, then we identify
$|\widetilde{k} \rangle = S^{-1/2} |k \rangle$ at
finite $g_2$.  From
the Hamiltonian $H = \Delta - J$
in the gauge theory basis we can construct the
basis-transformed operator
$\widetilde{H} = S^{1/2} H S^{-1/2}$, which we call the
string Hamiltonian.

In this note we prove an identity (Lemma 1, below)
satisfied by $\Sigma$ which implies
that
\eqn\aaa{
e^{g_2 \Sigma} H = H^\dagger e^{g_2 \Sigma}.
}
This ensures that the string Hamiltonian $\widetilde{H}$
is hermitian with respect to the
inner product \conjone. Obviously this is a basic requirement for identifying
$\widetilde{H}$ with the Hamiltonian of light-cone string
field theory in the plane-wave background.
Another consequence of the identity is that while $H$ in the
gauge theory has manifestly no contact terms according to \BeisertFF,
the string Hamiltonian $\widetilde{H}$ has
precisely one contact term at order $g_2^2$ (to first order in
$\lambda'$) \RoibanXR.
The contact term is therefore an artifact of the change of basis.
Finally, our formalism allows an analytic proof of
the fact that $\Delta - J$ is positive definite at one loop.
Of course, the supersymmetry algebra of the gauge theory requires
this to be true, but it is certainly not obvious
from the form of the generators given below, so it is satisfying
to see that this can be proven analytically even after one
has forgotten where the quantum mechanics comes from.

We conclude the paper with a discussion of some puzzles that
appear at higher loops, and some comments on the
S-matrix.
In particular, we propose to identify the S-matrix obtained from
this quantum mechanics with the S-matrix of string theory in the
plane wave background.

\newsec{Definition of the Quantum Mechanics}

In this section we review the definition the quantum
mechanics of BMN operators.
To this end we (1) explain the Hilbert space, (2) provide an
inner product, and (3) define some useful operators --- in particular,
the Hamiltonian.
The reader may choose to  think of this as an abstract quantum 
mechanical model.  However, some insight into its structure is
naturally gained by understanding precisely how these definitions
arise from the gauge theory, as will be explained in the next section.

\subsec{Hilbert space}

The Hilbert space is spanned by two kinds of basis vectors.
The first kind is
\eqn\typeone{
|n, r_0; r_1,\ldots,r_k\rangle, \qquad
n \in \IZ,~  k \ge 0, ~
~ r_0, r_i \in [0,1], ~ r_0 + \sum_{i=1}^k r_i = 1,
}
and the second kind is
\eqn\typetwo{
|s_1; s_2; r_1,\ldots,r_{k}\rangle, \qquad
k \ge 0, ~ s_1,s_2,r_i \in [0,1], ~ s_1 + s_2 + \sum_{i=1}^{k} r_i = 1.
}
The order of the $\{r_i\}$ is not significant.

\subsec{Inner product}

We define the inner product $\langle ~|~\rangle$ by
\eqn\ipdef{\eqalign{
\langle m, r'_0; r'_1,\ldots,r'_l|n, r_0;r_1,\ldots,r_k\rangle &= \delta_{kl}
\delta_{mn} \sum_{\pi \in S_k} \prod_{i=1}^k \delta(r'_i - r_{\pi(i)}),
\cr
\langle n, r_0; r_1,\ldots,r_l
| s_1; s_2; r_1,\ldots,r_k\rangle &= 0,\cr
\langle s'_1; s'_2;
r'_1,\ldots,r'_l| s_1;s_2; r_1,\ldots,r_k\rangle &=
\delta_{kl}
\delta(s_2' - s_2)
\sum_{\pi \in S_k} \prod_{i=1}^k \delta(r'_i - r_{\pi(i)}).
}}
We will use the symbol $\dagger$ to denote the adjoint with
respect to this inner product.

\subsec{Splitting-joining operator}

We now define a hermitian operator $\Sigma = \Sigma_+ + \Sigma_-$
which plays a central role.
The splitting operator $\Sigma_+$ is
defined by\foot{A technical comment:  note that
\eqn\simplify{
(-1)^n \sin(\pi n y) \sin(\pi n (1-y)) = - \sin^2(\pi n y) \qquad
{\rm for~} n \in \IZ.
}
However, we have not used this identity to simplify the first line
of $\Sigma_+$ because
\eqn\aaa{
{ (-1)^n \sin(\pi n y) \sin(\pi n(1-y)) \over n^2} \qquad
{\rm and} \qquad - {\sin^2(\pi n y) \over n^2}
}
disagree when $n=0$.  We will only use
\simplify\ when there is no such problem.
}
\eqn\spdef{\eqalign{
&\Sigma_+ | n, r_0;r_1,\ldots,r_k\rangle
=  {(-1)^n r_0^{3/2}\over \pi^2 n^2} \int_0^{r_0} ds
\  {\textstyle{\sin(\pi n { r_0 -s \over r_0}) \sin(\pi n {s \over r_0})}}
| r_0-s; s; r_1,\ldots,r_k\rangle
\cr
&\qquad
+\int_0^{r_0} dr_{k+1}
\sum_{m=-\infty}^\infty 
\sqrt{(r_0{-}r_{k+1})^3 r_{k+1} \over r_0}
{\sin^2(\pi n {r_0 - r_{k+1} \over r_0}) \over \pi^2
(m - n {r_0 - r_{k+1} \over r_0})^2
}
 |m, r_0{-}r_{k+1}; 
r_1,\ldots,r_{k+1}\rangle,
\cr
&\qquad+ \sum_{i=1}^k \int_0^{r_i} dr_{k+1} \sqrt{r_i (r_i - r_{k+1})
r_{k+1}} | n, r_0; r_1,\ldots,r_i - r_{k+1}, \ldots,r_{k+1}\rangle,\cr
&\Sigma_+ | s_1; s_2; r_1,\ldots,r_k\rangle
= \int_0^{s_1} dr_{k+1}
(s_1 - r_{k+1}) \sqrt{r_{k+1}}
|s_1 - r_{k+1}; s_2; r_1,\ldots,r_{k+1}\rangle
\cr
&\qquad + \int_0^{s_2} dr_{k+1}
(s_2 - r_{k+1}) \sqrt{r_{k+1}}
| s_1; s_2- r_{k+1}; r_1,\ldots,r_{k+1}\rangle
\cr
&\qquad + \sum_{i=1}^k \int_0^{r_i} dr_{k+1} \sqrt{
r_i (r_i - r_{k+1}) r_{k+1} } |s_1;s_2; r_1,\ldots,r_i-r_{k+1},\ldots,
r_{k+1}\rangle,
}}
and the joining operator
is $\Sigma_- = (\Sigma_+)^\dagger$.  (The formula is presented in
appendix C in order to avoid cluttering the text too much).
The nomenclature will become obvious in the next section;
for now one can note that  $\Sigma_+$ ($\Sigma_-$) 
increases (decreases) the number of `traces' by one.
It will also become obvious, for example, why $\Sigma_+$ never
takes a state of the second type into a state of the first type.

\subsec{Hamiltonian}

The free Hamiltonian is defined to be
\eqn\freeH{
H_0 | n, r_0; r_1, \ldots, r_k \rangle = {n^2 \over r_0^2}
|n,r_0; r_1,\ldots,r_k\rangle, \qquad
H_0 | s_1; s_2; r_1, \ldots, r_k \rangle=0.
}
It will prove convenient to introduce an operator $Q_0$ which is
a square root of the free Hamiltonian,
\eqn\aaa{
Q_0 | n, r_0; r_1, \ldots, r_k \rangle = {n \over r_0}
|n,r_0; r_1,\ldots,r_k\rangle, \qquad
Q_0 | s_1; s_2; r_1, \ldots, r_k \rangle=0.
}
Clearly both $H_0$ and $Q_0$ are hermitian.
The full Hamiltonian $H$ is given by
\eqn\aaa{
H = H_0 + g_2 V, \qquad V = H_+ + H_-,
}
where $g_2$ is the coupling constant and $H_\pm$ are the interaction
terms
\eqn\hpmdef{
H_\pm = Q_0 [ Q_0, \Sigma_\pm].
}
Using the definitions above, it is easy to see that
when acting on states of the first type,
\eqn\htypeone{
\eqalign{
&H_+ |n, r_0; r_1,\ldots,r_k\rangle
\cr
&\qquad
= \int_0^{r_0} dr_{k+1}
\sum_{m=-\infty}^\infty
\sqrt{r_{k+1} \over r_0(r_0 - r_{k+1})}
{ m \sin^2(\pi n {r_0-r_{k+1} \over r_0})
\over  \pi^2 (m-n
{r_0 - r_{k+1} \over r_0})} |m,r_0 - r_{k+1};r_1,\ldots,r_{k+1}\rangle,\cr
&H_- |n,r_0; r_1,\ldots,r_k\rangle
\cr
&\qquad= \sum_{i=1}^k \sum_{m=-\infty}^\infty
\sqrt{r_i \over r_0  (r_0 + r_i)}
{ m \sin^2(\pi m {r_0 \over r_0 + r_i})
\over \pi^2 (m - n {r_0 + r_i \over r_0})} |m, r_0+r_i;
r_1,\ldots,r_i\!\!\!\!\!\times,\ldots,r_k\rangle,
}}
while on states of the second type
\eqn\htypetwo{\eqalign{
H_+ |s_1; s_2 ; r_1,\ldots,r_k\rangle
&= 0,\cr
H_- | s_1; s_2 ; r_1,\ldots,r_k\rangle &=
- \sum_{m=-\infty}^\infty
{\sin^2(\pi m {s_1 \over s_1 + s_2})
 \over \pi^2 \sqrt{s_1 + s_2}}
|m, s_1 + s_2; r_1,\ldots,r_k\rangle.
}}
Note that $H_+$ is not the adjoint of $H_-$.
Therefore
$V$ (and hence the Hamiltonian
$H$) is not hermitian with respect
to the inner product $\langle ~ |~\rangle$ defined
in subsection 2.3.  At this point all we can say is
that
\eqn\vqqsigma{
V = Q_0 [Q_0, \Sigma], \qquad V^\dagger = [ \Sigma, Q_0 ] Q_0.
}

Let us note that the expression given in \htypeone\ 
differs from that of \BeisertFF\ in the following
inconsequential ways:  (1) their Hamiltonian has an additional
factor of ${4 \pi^2}$ (which
can be absorbed into $\lambda'$), (2) the expressions \htypeone\ have
slightly different factors in the square root, owing to a slightly
different definition of the states (see next section), and (3)
the arguments of the $\sin^2$ functions are slightly different,
but equivalent (since $m$ and $n$ are integers).
Finally, we remark that the authors of \BeisertFF\ 
had no need for the expressions \htypetwo\ because they
focused on diagonalizing $H$ within the subspace of the first
kind of state.  Since acting with $H$ will never produce
states of the second type, it is consistent for their purposes
to completely disregard the second component of the Hilbert space.

\newsec{Relation to Gauge Theory}

Here we summarize the relation between the definitions in the
previous section and the BMN limit of the gauge theory.
This section is provided for cultural enrichment; those readers 
content to study the structure of the quantum mechanics for
its own sake may proceed to section 4.

\subsec{Hilbert space}

Recall that ${\cal{N}} = 4$ SU($N$) super Yang-Mills theory
has 6 real scalar fields $\phi_i$.  The BMN operators can
be constructed from three orthogonal
complex combinations, which can be taken to be
\eqn\aaa{
\phi = {1 \over \sqrt{2}} (\phi_1 + i \phi_2), \qquad
\psi = {1 \over \sqrt{2}} (\phi_3 + i \phi_4), \qquad
Z = {1 \over \sqrt{2}} (\phi_5 + i \phi_6).
}
The state-operator identification is
then
\eqn\opmap{\eqalign{
|n,r_0; r_1,\ldots,r_k\rangle&\leftrightarrow
\int_0^{r_0} dx\ {e^{2 \pi i n x/r_0}
\over \sqrt{r_0 r_1 \cdots r_k}}
\Tr(\phi Z^{Jx} \psi Z^{J(r_0 - x)})
\Tr(Z^{J_1}) 
\cdots \Tr(Z^{J_k}),
\cr
{| s_1; s_2; r_1,\ldots,r_k\rangle}
&\leftrightarrow {1 \over \sqrt{r_1 \cdots r_k}}
\Tr( \phi Z^{J s_1} ) \Tr(\psi Z^{J s_2})
\Tr(Z^{J_1}) \cdots \Tr(Z^{J_k}),
}}
where $r_i = J_i/J$.
We will use $O_{|a\rangle}$ to denote the operator corresponding
to the state $|a\rangle$.

\subsec{Inner product}

The gauge theory inner product is related by the state-operator mapping to
the two-point function in the free ($g_{\rm YM} = 0$) theory
according to the formula
\eqn\fullip{
\left.
\langle {\overline{O}}_{|1\rangle}
(0) {{O}}_{|2\rangle}(x) \rangle \right|_{\rm free} =
{J^{-1} N^{J+2} \over (4 \pi^2 x^2)^{J+2}} \langle 1 | 2 \rangle_{g_2}.
}
This formula may be viewed as the definition of
$\langle 1|2\rangle_{g_2}$.

The factor of $J^{-1}$ in \fullip\ can be motivated by checking
this relation in the $g_2 \to 0$ limit.
For operators of the first type one gets a factor of
$J^{k+1}$ (at large $J$)
from contracting the fields, a factor of $J^{-2}$
from converting the integrals in \opmap\ to sums, and
a factor of $J^{-k}$ from converting $k$ Kronecker delta-functions
to the continuous delta functions in the first line of \ipdef.
For operators of the second type one gets a factor
of $J^k$ from contracting the $Z$ fields and a factor
of $J^{-k-1}$ from converting $k+1$ Kronecker delta-functions.

\subsec{Splitting-joining operator}

The inner product we defined in subsection 2.2 corresponds
to the gauge theory inner product \fullip\ only at $g_2 = 0$.
The splitting-joining operator $\Sigma$ which we
defined above gives the first order $g_2$
correction to the inner product according to the formula
\eqn\sigmadef{
\left.
\langle {\overline{O}}_{|1\rangle}(0) {{O}}_{|2\rangle}(x) \rangle
\right|_{{\rm free}, {\cal{O}}(g_2)}
=g_2
{J^{-1} N^{J+2} \over (4 \pi^2 x^2)^{J+2}} \langle 1 | \Sigma | 2 \rangle.
}
Equivalently: matrix elements of $\Sigma$ may be computed
in the gauge theory by calculating free, planar contractions between
$k$-trace and $k+1$-trace BMN operators.

This operator $\Sigma$ has appeared in at least three different
but equivalent guises in the
plane wave literature:

(1) It encodes free, planar three-point functions of BMN operators 
\refs{\ConstableHW,\BeisertBB,\ConstableVQ}.

(2) It is the permutation operator in the discretized string
theory (bit model) \refs{\VerlindeIG, \VamanKA}.

(3) In light-cone string field theory, $\Sigma$ is the
three-string vertex without prefactor \refs{\SpradlinAR, \SpradlinRV}.

\subsec{Hamiltonian}

The Hamiltonian of the quantum mechanical model corresponds
in the gauge theory to $\Delta - J$, where $\Delta$ is the
dilatation operator.
At $g_2 = 0$ it is known that
\eqn\aaa{\eqalign{
(\Delta - J) |n,r_0;r_1,\ldots,r_k\rangle&=
2 \sqrt{1 + \lambda' {n^2/r_0^2}}
|n, r_0; r_1,\ldots,r_k\rangle,\cr
(\Delta-J) |s_1;s_2;r_1,\ldots,r_k\rangle&= 0,
}}
and the expressions given for $H$ in section 2 encapsulate the one-loop
(i.e. ${\cal{O}}(\lambda')$) contribution to $\Delta - J$.
The interaction terms $H_+ + H_-$ encode the first
order (in $g_2$) elements of the one-loop
anomalous dimension mixing matrix.

\newsec{The String Basis}

In this section we investigate some properties of
the  quantum mechanics.
Of paramount importance is
the relation
\eqn\cool{
S H = H^\dagger S, \qquad S = e^{g_2 \Sigma},
}
to be proven below.
Let us explain the significance of this result.  We
saw in section 2 that the Hamiltonian $H$ is not hermitian
with respect to the inner product $\langle ~|~\rangle$.  However, this is
no cause for concern: while $H$ is manifestly hermitian
with respect to the gauge theory inner product $\langle ~
| ~\rangle_{g_2}$ of \fullip, there is no reason for it to
be hermitian with respect to $\langle ~ | ~\rangle$ since the
two inner products agree only at $g_2 = 0$!

It is believed that the gauge theory inner product at finite $g_2$
(defined by \fullip)
is given by the simple formula
\eqn\newip{
\langle  1|  2 \rangle_{g_2} = \langle 1 | S | 2 \rangle.
}
The relation \cool\ guarantees that $H$ is hermitian with
respect to \newip.  This provides a consistency check
on the conjecture that \newip\ is actually the correct gauge theory
inner product at finite $g_2$.

\subsec{Some identities}

 From the definition \vqqsigma\ we immediately obtain the relations
\eqn\bbb{
[H_0, \Sigma] = V - V^\dagger, \qquad [ Q_0, [Q_0, \Sigma]] = V + V^\dagger.
}
Next we present

\noindent
{\bf Lemma 1.}
\eqn\ccc{
[ \Sigma, [ \Sigma, Q_0 ]] = 0.
}
The lengthy proof of this result is given in appendix A.
It would be interesting to understand the gauge theory origin of this lemma,
which might be possible to derive as a consequence of supersymmetry.

Some immediate consequences of Lemma 1 which follow directly from
the definition \vqqsigma\ include
\eqn\eee{
[ \Sigma, V + V^\dagger ]  = 0, \qquad [ \Sigma, [ \Sigma, V]] = 0.
}
Next consider the formula
\eqn\ggg{
e^A B e^{-A} = B + [ A,B] + {1 \over 2!} [ A, [A,B]]
+ {1 \over 3!} [ A, [A, [A,B]]] + \cdots
}
which leads to
\eqn\fff{\eqalign{
e^{\lambda g_2 \Sigma} H_0 e^{- \lambda g_2 \Sigma}
&= H_0 + g_2
\lambda ( V^\dagger - V) + {g_2^2 \over 2}
\lambda^2 [ \Sigma, V^\dagger - V],\cr
e^{\lambda g_2 \Sigma} g_2 V e^{- \lambda g_2 \Sigma}
&= g_2 V + {g_2^2 \over 2} \lambda [ \Sigma, V - V^\dagger],
}}
for an arbitrary parameter $\lambda$.
Remarkably, all higher order terms vanish as a consequence 
of \eee.
Combining \fff\ gives
\eqn\remark{
e^{\lambda g_2 \Sigma} H e^{-\lambda g_2 \Sigma}
= H - g_2 \lambda (V - V^\dagger)
+ {g_2^2 \over 2} (\lambda - \lambda^2)
[ \Sigma, V - V^\dagger].
}
For $\lambda = 1$
we find
\eqn\aaa{
S H S^{-1} = H^\dagger,
}
thereby establishing the desired relation \cool.

\subsec{Hamiltonian in the string basis}

The basis which diagonalizes the inner product \newip\ is 
\eqn\aaa{
| \widetilde{a} \rangle = S^{-1/2} |a \rangle.
}
This basis has been identified as the basis of string states
in the light-cone string field theory:
if $|k\rangle$
corresponds to an operator with precisely
$k$ traces, then $S^{-1/2} |k\rangle$
corresponds to a state of precisely
$k$ strings.

It is convenient to
define an operator $\widetilde{H}$ whose matrix elements in
the gauge theory basis are the same as the matrix elements of
the Hamiltonian $H$ in the string basis:
\eqn\aaa{
\langle \widetilde{a} | H | \widetilde{b} \rangle_{g_2}
= \langle a | \widetilde{H} |  b\rangle.
}
We will call $\widetilde{H}$ the `string Hamiltonian',
since its matrix elements are related to matrix elements
of the light-cone string field theory Hamiltonian.
Clearly, $\widetilde{H}$ is given by
\eqn\aaa{
\widetilde{H} = S^{1/2} H S^{-1/2}.
}
Setting $\lambda = \ha$ in
\remark\ leads to
\eqn\wth{
\widetilde{H} = H_0 + {g_2 \over 2} ( V + V^\dagger) +
{g_2^2 \over 8} [ \Sigma, V - V^\dagger].
}
Naturally, $\widetilde{H}$ is manifestly hermitian with respect
to the inner product $\langle ~|~\rangle$.
It is remarkable that whereas the Hamiltonian
in the gauge theory basis has no contact terms,
the basis transformation introduces
a single ${\cal{O}}(g_2^2)$ contact
term.

\subsec{The `supercharge'}

Remarkably, it is easy to see that the string Hamiltonian $\widetilde{H}$
is a perfect square!  In particular,
\eqn\aaa{
\widetilde{H} = Q^\dagger Q, \qquad Q = Q_0 + {g_2 \over 2} [Q_0,\Sigma]
= S^{-1/2} Q_0 S^{1/2}.
}
The order $g_2$ term in \wth\ works out due to \bbb, while the
$g_2^2$ term follows from \vqqsigma\ and
\eqn\aaa{\eqalign{
[ \Sigma, Q_0 [ Q_0, \Sigma]] - [ \Sigma, [ \Sigma, Q_0] Q_0] -
2 [ \Sigma, Q_0] [ Q_0, \Sigma] = 0,
}}
which is 
a consequence of \ccc.
Note that it follows from Lemma 1 that
\eqn\aaa{
[\Sigma, [\Sigma, Q]]=0.
}
Of course, the ${\cal{N}}=4$ supersymmetry algebra
requires that $\Delta - J$ is positive definite.  However, it is
nice to see that this can be analytically proven from
the expressions
\htypeone, where this fact is not at all obvious.

One interesting open problem is to supersymmetrize this
quantum mechanics by including fermionic impurities, in which
case the appropriate generalization of $Q$ would
become an honest fermionic supercharge.  It would be very
interesting to see whether the fermionic extension of this
quantum mechanics proceeds as in the string bit model \VamanKA, or whether
the fermionic completion has a different flavor here.

\subsec{The string field theory `prefactor'}

Let us consider order $g_2$ matrix elements of $\widetilde{H}$
between two energy eigenstates --- actually, two eigenstates
of $Q_0$ with eigenvalues $\sqrt{E_1}, \sqrt{E_2}$.  From
\wth\ and \bbb\ we immediately have
\eqn\aaa{
\langle 1 | \widetilde{H} |2 \rangle = {g_2 \over 2}
\langle 1 | [Q_0, [ Q_0, \Sigma]]  | 2\rangle
= {g_2 \over 2} ( \sqrt{E_1} - \sqrt{E_2})^2 \langle
1 | \Sigma | 2 \rangle.
}
This formula equates matrix elements of the string
field theory Hamiltonian $\widetilde{H}$ to
three-point functions $\langle 1 | \Sigma | 2 \rangle$ in
the gauge theory, with a certain `prefactor'.
In the earliest literature on the subject, the prefactor
was erroneously believed to be $E_1 - E_2$ instead of
$\ha ( \sqrt{E_1} - \sqrt{E_2})^2$.
It would be very interesting to understand the generalization
of this formula for more complicated processes, in particular those
involving more than two impurities.

\newsec{Discussion and Speculations}

\subsec{A puzzle at two loops}

A study of the gauge theory at two-loops has recently been presented in
\BeisertTQ.  The results led the authors to the conjecture
that the Hamiltonian is given to all loops by
\eqn\aaa{
H_{\rm full} = 2 \sqrt{1 + \lambda' H}
}
(with firm calculational
support only up to and including ${\cal{O}}(\lambda'^2)$),
where $H$ on the right-hand side is the one-loop Hamiltonian
studied in the previous sections.
The Hamiltonian in the string basis would then be
\eqn\aaa{
\widetilde{H}_{\rm full} = S^{1/2} H_{\rm full} S^{-1/2}
= 2 \sqrt{1 + \lambda' \widetilde{H}}.
}
It is not hard to derive from this a formula for the
order $g_2$ contribution to the following matrix element,
to all orders in $\lambda'$:
\eqn\bada{\eqalign{
{ \langle n, 1| \widetilde{H}_{\rm full} | m, y; 1-y\rangle
\over \langle n,1| \widetilde{H} | m,y; 1-y\rangle}
&= 
{ \sqrt{ 1 + \lambda' n^2} - \sqrt{1 + \lambda' m^2/y^2}
\over {1 \over 2} \lambda' (n^2 - m^2/y^2)}
 + {\cal{O}}(g_2)\cr
&= 
\left[ 1 - {1 \over 4}\lambda' (n^2 + m^2/y^2) + {\cal{O}}(\lambda'^2)
\right] + {\cal{O}}(g_2).
}}

On the other hand,
this particular matrix element has also been studied extensively
on the string field theory side of the BMN correspondence.  In
\HeZU\ it was shown that to all orders in $\lambda'$ perturbation theory,
\eqn\badb{\eqalign{
{ \langle n, 1| P^-_{{\rm all~orders~in~} \lambda'} | m,y;1-y\rangle
\over \langle n,1| P^-_{{\rm first~order~in~}\lambda'} |m,y;1-y\rangle}&=
{1 \over \sqrt{1 + \lambda' n^2} \sqrt{1 +\lambda'
m^2/y^2}}\cr
&= 1 - {1 \over 2} \lambda' (n^2 + m^2/y^2) + {\cal{O}}(\lambda'^2).
}}

The results \bada\ and \badb\ disagree by a factor of 2 at two loops
(where the calculations of
\BeisertTQ\ are firm), and they disagree even more strongly
at higher loops (where \BeisertTQ\ merely conjectured).
In particular, it is impossible to write down any function
just of $\widetilde{H}$ that reproduces the result \badb\ to all
orders in $\lambda'$.
It is possible that this discrepancy involves
an unallowed exchange of the order of limits between gauge theory
and string field theory.  This problem has manifested itself in
\KlebanovMP, for example (where a ``renormalization'' by
a finite amount occurs below (2.14)), and in \RoibanXR,
regarding the issue of intermediate states which do not conserve
the number of impurities.  It is an interesting open problem to
understand precisely which observables we might expect to be
able to study perturbatively on both sides of the duality.

\subsec{The S-matrix}

What is it that we would most like to know about the quantum 
mechanics studied in this paper; i.e.~what is the ultimate goal
of this course of research?
We propose that the goal should be to calculate the
non-relativistic S-matrix obtained from this quantum mechanics,
which we identify with the S-matrix of
string theory \BakKU\ in the plane wave background, after the
appropriate basis transformation.\foot{It is important
to note that in general an S-matrix depends on how one chooses
to divide the full Hamiltonian into a `free' part and an
`interacting' part.  In our case, the basis transformation
$S^{1/2}$ does not commute with this division:  the
free and interacting parts of $\widetilde{H}$ are not respectively
the transforms of the free and interacting parts of $H$.
Therefore the S-matrices obtained from $H$ and $\widetilde{H}$
are not unitarily related to each other (although it
would be intriguing to see if there were some other, more complicated
relationship between them).}

Much of the literature on this subject has focused
(quite successfully) on comparing matrix elements of $\widetilde{H}$
to matrix elements of the Hamiltonian $P^-$ of light-cone
string field theory in the plane wave background.
The S-matrix proposal subsumes all of the supporting evidence (since it
is a weaker proposal) and simultaneously satisfies the
ardent skeptic who points out that only the S-matrix is
a good
observable in string theory (as matrix elements of the light-cone
Hamiltonian are not coordinate invariant).

The authors of \BeisertFF\ set out to find
the spectrum of $H$ (which is identical to the
spectrum of $\widetilde{H}$).
This would of course be very useful to know,
but this might be a very difficult task in practice.  The BMN
correspondence suggests that $H$ should be related somehow to
the Hamiltonian of an interacting string theory, which likely
has an exceedingly complicated spectrum! Indeed the authors
of \BeisertFF\ encountered technical difficulties at genus two
due to overlapping continuum states.
However, one rarely studies string theory (or any
quantum field theory) by attempting to diagonalize
the Hamiltonian.  Instead, the goal is usually to calculate
the S-matrix.

Let us now recall some non-relativistic scattering theory.
We write the string Hamiltonian as
\eqn\aaa{
\widetilde{H} = H_0 + W, \qquad W = {g_2 \over 2} (V + V^\dagger)
+  {g_2^2 \over 8} [ \Sigma, V - V^\dagger].
}
Then the transition matrix $T(z)$ can be obtained from
the Born series
\eqn\aaa{
T(z) = W + W G_0(z) W + W G_0(z) W G_0(z) W + \cdots,
}
where $G_0(z) = (z - H_0)^{-1}$ is the free propagator.
As a function of the complex variable $z$,
the operator $T(z)$ should have poles
at the bound states of $H$ and branch cuts along the continuous
spectrum of $H$.
The S-matrix is then given by
\eqn\aaa{
\langle 1 | S  | 2\rangle =  \langle 1|
1 - 2 \pi i \delta(E_1 - E_2) T(E_1 + i \epsilon) |2\rangle,
}
when $H_0 | i\rangle = E_i |i\rangle$.

It is now possible (though technically complicated) to
calculate the S-matrix to any desired order in $g_2$.
The divergences encountered in \BeisertFF\ at genus two would
also occur if one tried to diagonalize the Hamiltonian
of string theory in the plane wave background.  According to
our proposal, they should be regulated with a $+i
\epsilon$ prescription (instead of a principal value) and
interpreted as the usual branch
cuts one finds in transition amplitudes when there is a continuum
of intermediate scattering states.   

\bigskip

\noindent
{\bf Acknowledgements}
\smallskip
It is a pleasure to thank D. Freedman, U. Gursoy, N. Itzhaki,
R. Roiban,
D. Vaman and H. Verlinde for helpful discussions.
The work of M.S. was supported in part by DOE
grant DE-FG02-91ER40671, and that of A.V. was supported in part by
by the National Science Foundation under
Grant No.~PHY99-07949.

\appendix{A}{Proof of Lemma 1}

By separating $\Sigma = \Sigma_+ + \Sigma_-$ and
defining $P_\pm = [Q_0, \Sigma_\pm]$,  we can express
\ccc\ as three relations which must separately be satisfied:
\eqn\relations{\eqalign{
(1) & \qquad [ \Sigma_-, P_- ] = 0,\cr
(2) & \qquad
[ \Sigma_-, P_+ ] + [ \Sigma_+, P_- ] = 0,\cr
(3) & \qquad
[ \Sigma_+, P_+] = 0.
}}
Of course, (3) is is equivalent to the adjoint of (1) so it will suffice
to check only the relations (1) and (2).  From
the definitions in section 2 we have
\eqn\aaa{\eqalign{
&P_+ |n, r_0;r_1,\ldots,r_k\rangle
= 
\sqrt{r_0}  \int_0^{r_0} ds {\sin^2(\pi n {s \over r_0})
\over \pi^2 n}
 |r_0-s; s; r_1,\ldots,r_k\rangle\cr
&\qquad+ \int_0^{r_0} dr_{k+1} \sum_{m=-\infty}^\infty
\sqrt{(r_0 - r_{k+1}) r_{k+1} \over
r_0}
{\sin^2(\pi n {r_0 - r_{k+1} \over r_0})
\over \pi^2 (m - n {r_0 - r_{k+1} \over r_0})}
|m, r_0 - r_{k+1};r_1,\ldots,r_{k+1}\rangle,\cr
&P_- |n, r_0; r_1,\ldots,r_k\rangle\cr
&\qquad=\sum_{i=1}^{k} 
\sum_{m=-\infty}^\infty
\sqrt{(r_0 + r_i) r_i \over r_0}
{\sin^2(\pi m {r_0 \over r_0 + r_i})
\over \pi^2 (m - n {r_0 + r_i\over r_0})}
|m, r_0+r_i; r_1,\ldots,r_i\!\!\!\!\!\times,\ldots,r_k\rangle
}}
on states of type one and
\eqn\aaa{\eqalign{
P_+ |s_1; s_2; r_1,\ldots,r_k\rangle &= 0,\cr
P_- |s_1;s_2;r_1,\ldots,r_k\rangle &=
- \sqrt{s_1 + s_2} \sum_{m=-\infty}^\infty
{ \sin^2(\pi m {s_1 \over s_1 + s_2}) \over \pi^2 m}
| m, s_1 + s_2; r_1,\ldots,r_k\rangle
}}
on states of the second type.
We now consider separately the relations \relations\ on the two
kinds of states.

\subsec{Relation (1) on states of the first type}

Acting with $\Sigma_- P_-$ on a state of type one gives
\eqn\atwo{
\eqalign{
&\Sigma_- P_- | n, r_0; r_1,\ldots,r_k \rangle
= \sum_{i \ne j} \sum_{m=-\infty}^\infty
A_{ijm}
|m, r_0 + r_i + r_j; r_1, \ldots, r_i\!\!\!\!\!\times,\ldots,
r_j\!\!\!\!\!\times,\ldots,r_k\rangle
\cr
&
+ \sum_{i=1}^k \sum_{m=-\infty}^\infty \sqrt{(r_0 + r_i) r_i \over
r_0} {\sin^2(\pi m {r_0 \over r_0 + r_i}) \over
\pi^2 (m - n {r_0 + r_i \over r_0})}
\cr
&\qquad\times
\sum_{j \ne i} \sum_{l \ne j, i}
\ha \sqrt{r_j r_l(r_j+r_l)}
| m, r_0 + r_i; r_1,\ldots,r_i\!\!\!\!\!\times,\ldots
r_j\!\!\!\!\!\times,\ldots,r_l\!\!\!\!\!\times,\ldots
r_k,r_j+r_l\rangle,
}}
where the coefficient $A$ is given by
\eqn\xlsss{
A_{ijm} =\sum_{p=-\infty}^\infty
 \sqrt{  r_i (r_0 + r_i + r_j)^3 r_j \over
r_0 } { \sin^2(\pi p {r_0 \over r_0 + r_i})
\over \pi^2 (p - n {r_0 + r_i \over r_0})}
{\sin^2(\pi m {r_0 + r_i \over r_0 + r_i + r_j} )
\over \pi^2(m - p {r_0 + r_i + r_j \over r_0 + r_i})^2}.
}
Similarly, acting with $P_- \Sigma_-$ gives
\eqn\athree{
\eqalign{
&P_- \Sigma_- |n,r_0;r_1,\ldots,r_k\rangle
= \sum_{i \ne j} \sum_{m=-\infty}^\infty B_{ijm} |m,r_0+r_i+r_j;
r_1,\ldots,r_i\!\!\!\!\!,\ldots,r_j\!\!\!\!\!,\ldots,r_k\rangle\cr
&\qquad\qquad\qquad + {1 \over 2} \sum_{j\ne l}
\sqrt{r_j r_l (r_j + r_l)}
P_- |n, r_0; r_1,\ldots,r_j\!\!\!\!\!\times,\ldots
r_l\!\!\!\!\!\times,\ldots,r_k,r_j+r_l\rangle,
}}
with coefficient
\eqn\aaa{
B_{ijm} =
\sum_{p=-\infty}^\infty \sqrt{(r_0 + r_i)^2 r_i (r_0 + r_i + r_j)
r_j \over r_0 }
{ \sin^2(\pi p {r_0 \over r_0 + r_i}) \over\pi^2 (p - n
{ r_0 + r_i \over r_0})^2}
{ \sin^2(\pi m {r_0 + r_i \over r_0 + r_i + r_j})
\over \pi^2 (m - p {r_0+r_i+r_j \over r_0+r_i})}.
}
Let us first study the $A_{ijm}$ and $B_{ijm}$ terms.
If we define
\eqn\aaa{
y = {r_0 \over r_0 + r_i}, \qquad a = n/y, \qquad b = m {r_0 + r_i \over
r_0 + r_i + r_j}
}
then we can combine 
$A$ and $B$ into the formula
\eqn\aaa{
\eqalign{
&A_{ijm} -  B_{ijm} = {1 \over
\pi^4 }\sqrt{r_i r_j \over r_0 (r_0 + r_i + r_j)}
\cr
&\qquad \times\left[
{(r_0 + r_i)^2 } {\textstyle{
\sin^2(\pi m {r_0 + r_i \over r_0 + r_i + r_j})
}}
\sum_{p=-\infty}^\infty {\sin^2(\pi p y) \over (p-a) (p-b)}
\left( {1 \over p-a} + {1 \over p-b}\right)
\right].
}}
The sum over $p$ can be performed with the help of (B.4),
and we find after much simplification
\eqn\afour{
A_{(ij)m} - B_{(ij)m}
= {r_i+r_j \over 2 \pi^2} \sqrt{r_i r_j (r_0 + r_i + r_j)\over r_0}
{{\sin^2(\pi m {r_0 \over r_0 + r_i + r_j})} \over
m - n {r_0 + r_i + r_j \over r_0}}.
}
Notice that we have symmetrized in the $i$ and $j$ indices.
This step is allowed (indeed, forced upon us) because
of the manifest $i \leftrightarrow j$ symmetry of the states these
coefficients multiply.

Now consider the second term in \atwo\ and the second term in \athree.
It is easy to see that they are essentially the same.  However, the
latter has an additional term when $P_-$ acts on the last trace
($r_j + r_l$), giving the term
\eqn\aaa{\eqalign{
&\cdots + \ha \sum_{j \ne l} \sqrt{r_j r_l(r_j + r_l)}
\sum_{m=-\infty}^\infty \sqrt{(r_0 + r_j + r_l)(r_j + r_l)
\over r_0}\cr
&\qquad\qquad\qquad\times {\sin^2(\pi m {r_0 \over r_0 + r_j+r_l})
\over \pi^2 (m - n {r_0+r_j+r_l \over r_0})}
|m,r_0+r_j+r_l; r_1,\ldots,r_j\!\!\!\!\!\times,\ldots,
r_l\!\!\!\!\!\times,\ldots,r_k\rangle.
}}
This additional term is precisely the same as the one whose coefficient
is given by \afour!  This completes the proof that
\eqn\aaa{
[ \Sigma_-, P_-] |n,r_0;r_1,\ldots,r_k\rangle = 0.
}

\subsec{Relation (1) on states of the second type}

Acting with $\Sigma_- P_-$ on a state of type two gives
something of the form
\eqn\formaaa{
\eqalign{
&\Sigma_- P_- | s_1;s_2; r_1,\ldots,r_k\rangle
= \sum_{i=1}^k \sum_{m=-\infty}^\infty
C_{im}
|m,s_1+s_2+r_i; r_1,\ldots,r_i\!\!\!\!\!\times,\ldots,r_k\rangle
\cr
&\qquad - \sqrt{s_1 + s_2} \sum_{m=-\infty}^\infty
{\sin^2(\pi m {s_1 \over s_1 + s_2}) \over
\pi^2 m}\cr
&\qquad\qquad\times {1 \over 2} \sum_{i \ne j}
\sqrt{r_i r_j(r_i+r_j)} |m,s_1+s_2;
r_1,\ldots,r_i\!\!\!\!\!\times,\ldots,r_j\!\!\!\!\!\times,\ldots,r_k,r_i
+r_j\rangle.
}}
with
\eqn\aaa{
\eqalign{
C_{im} &=
- \sqrt{ (s_1 + s_2 + r_i)^3 r_i}
\sum_{p=-\infty}^\infty
{ \sin^2(\pi p {s_1 \over s_1 + s_2})
\over \pi^2 p} {\sin^2(\pi m {s_1 + s_2 \over s_1 + s_2 + r_i})
\over \pi^2 (m-p {s_1 + s_2 + r_i \over s_1 + s_2})^2}.
}}
Similarly, from $P_- \Sigma_-$ we get an expression similar
to \formaaa.  The second term is identical, but the first term has
the coefficient
\eqn\aaa{\eqalign{
D_{im} &=
 \sqrt{(s_1 + s_2 + r_i) r_i}\Bigg[
- s_1 { \sin^2(\pi m {s_2  \over s_1 + s_2 + r_i})
\over \pi^2 m}
- s_2 { \sin^2(\pi m { s_1 \over s_1 + s_2 + r_i}) \over \pi^2 m}
\cr
&~~~~~
+(s_1 + s_2)
\sum_{p=-\infty}^\infty
{ (-1)^p \sin(\pi p {s_1 \over s_1 + s_2}) \sin(\pi p {s_2 \over s_1 +
s_2}) \over \pi^2 p^2}
{\sin^2(\pi m {s_1 + s_2 \over s_1 + s_2 + r_i})
\over \pi^2 (m - p {s_1 + s_2 + r_i \over s_1 + s_2})}
\Bigg].
}}
Applying the formulas in appendix B leads eventually to
\eqn\aaa{
C_{im} = D_{im},
}
thereby establishing that
\eqn\aaa{
[ \Sigma_-, P_-] |s_1;s_2;r_1,\ldots,r_k\rangle = 0.
}

\subsec{Relation (2) on states of the second type}

Since $P_+$ annihilates states of the second type, we have simply
\eqn\formbbb{\eqalign{
&[ \Sigma_-, P_+] |s_1; s_2;r_1,\ldots,r_k\rangle=
\int_0^{s_1 + s_2} dr_{k+1} \sum_{m=-\infty}^\infty
E_m | m; s_1 + s_2 - r_{k+1}; r_1,\ldots,r_{k+1}\rangle
\cr
&+ (s_1 + s_2)^2\sum_{m=-\infty}^\infty
{\sin^2(\pi m {s_1  \over s_1 + s_2}) \over \pi^2 m^2}
\int_0^{s_1 + s_2} ds {\sin^2(\pi m {s \over s_1 + s_2})\over \pi^2 m}
|s_1 + s_2-s;s;r_1,\ldots,r_k\rangle.
}}
The second term vanishes because the summand is odd in $m$, and
in the first term appears the coefficient
\eqn\aaa{
E_m = (s_1 + s_2)
\sqrt{(s_1 + s_2 - r_{k+1}) r_{k+1}}
\sum_{p=-\infty}^\infty {\sin^2(\pi p {s_1 \over s_1 + s_2})
\over \pi^2 p^2} {\sin^2(\pi p {s_1 + s_2 - r_{k+1} \over s_1 + s_2})
\over \pi^2 ( m - p {s_1 + s_2 - r_{k+1} \over s_1 + s_2})}.
}
Next  we consider
$[ \Sigma_+, P_-] |s_1;s_2; r_1,\ldots,r_k\rangle$.
Again there appears a term involving states of the second kind
which vanishes due to an odd summand.  The remaining terms of
the first type take the same form as in \formbbb, but with a much
more complicated coefficient
\eqn\aaa{
\eqalign{
F_m &= - \sqrt{ (s_1 + s_2 - r_{k+1})^3 r_{k+1}}
\sum_{p=-\infty}^\infty { \sin^2(\pi p {s_1 \over s_1 + s_2})
\over \pi^2 p}
 {\sin^2(\pi p {s_1 + s_2 - r_{k+1} \over s_1 + s_2})
\over \pi^2 (m - p {s_1 + s_2 - r_{k+1} \over s_1 + s_2})^2}\cr
&+ (s_1 - r_{k+1}) \sqrt{ (s_1 + s_2 - r_{k+1}) r_{k+1}}
\ {\sin^2(\pi m {s_1 - r_{k+1} \over
s_1 + s_2 - r_{k+1}}) \over \pi^2 m} \theta(s_1 - r_{k+1})\cr
&+ 
(s_2 - r_{k+1}) \sqrt{(s_1 + s_2 - r_{k+1}) r_{k+1}}
{\sin^2(\pi m {s_2 - r_{k+1} \over s_1 + s_2 - r_{k+1}}) \over\pi^2 m}
\theta(s_2 - r_{k+1}).
}}
We are attempting to show that
\eqn\aaa{
\left(
[ \Sigma_-, P_+] + [ \Sigma_+, P_-]\right) |s_1;s_2;r_1,\ldots,r_k\rangle = 0.
}
This requires $E_m + F_m = 0$, which can be written as
\eqn\hard{\eqalign{
& 
\sum_{p=-\infty}^\infty { \sin^2(\pi p {s_1 \over s_1 + s_2})
\sin^2(\pi p {s_1 + s_2 - r_{k+1} \over s_1 + s_2}) \over
p (p-a)} \left[ {1 \over p} + {1 \over p-a} \right]
=
{\pi^2 (s_1 + s_2 - r_{k+1})\over m(s_1 + s_2)^2}\cr
&\qquad\qquad\qquad\times \left[
(s_1 - r_{k+1})
\ {\textstyle{\sin^2(\pi m {s_1 - r_{k+1} \over
s_1 + s_2 - r_{k+1}})}} \theta(s_1 - r_{k+1})
+ (s_1 \leftrightarrow s_2)\right],
}}
where $a = m {s_1 + s_2 \over s_1 + s_2 - r_{k+1}}$.
This highly nontrivial identity is a consequence of (B.6), but requires
some  explanation since the constraint $1 \ge y_1 + y_2 \ge y_1 \ge y_2
\ge 0$ in (B.6) is crucial.
The integral in \formbbb\ splits into three regions.  In each
region a different choice of $y_1$ and $y_2$ is necessary in order
to satisfy the constraint.  Taking $s_2 \ge s_1$ without loss of
generality, the appropriate choices are
\eqn\aaa{
\eqalign{
r_{k+1} \in [0, s_1] &: \qquad y_1 = {s_1 \over s_1 + s_2}, 
\qquad y_2 = {r_{k+1} \over s_1 + s_2},\cr
r_{k+1} \in [s_1, s_2] &: \qquad y_1 = {s_1 + s_2 - r_{k+1} \over s_1
+ s_2}, \qquad y_2 = {s_1 \over s_1 + s_2},\cr
r_{k+1} \in [s_2, s_1+s_2] &:\qquad y_1 = {s_1 \over s_1 + s_2},
\qquad y_2 = {s_1 + s_2 - r_{k+1} \over s_1 + s_2}.
}}
After much simplification, we find in each case that (B.6)
reproduces the right-hand side of \hard.

\subsec{Relation (2) on states of the first type}

When we act with $[\Sigma_-, P_+] + [\Sigma_+, P_-]$ on a state of the first
type
we get many terms.  Let us first collect four terms
$T_1,\ldots,T_4$ which are of the
second type.  From $\Sigma_- P_+$ we get
\eqn\bhf{\eqalign{
&T_1=
{\sqrt{r_0} \over \pi^2 n} \int_0^{r_0} ds\ {\textstyle{\sin^2(\pi n {s \over
r_0})}} \sum_{i=1}^k
\Bigg[ (r_0 - s)\sqrt{r_i} |r_0-s+r_i;s; r_1,\ldots,r_i\!\!\!\!\!\times,\ldots,
r_k\rangle
\cr
&\qquad
\qquad\qquad
\qquad\qquad
\qquad\qquad
\qquad
+ s \sqrt{r_i}
|r_0 - s; s + r_i;  r_1,\ldots,r_i\!\!\!\!\!\times,\ldots,r_k\rangle
\Bigg]
}}
and
\eqn\aaa{\eqalign{
& T_2= \sqrt{r_0} \int_0^{r_0} ds {\sin^2(\pi n {s \over r_0})
\over \pi^2 n} 
\cr
&\qquad\qquad\qquad\times
{1 \over 2} \sum_{i \ne j}
\sqrt{r_i r_j (r_i + r_j)} |r_0 - s; s; r_1,\ldots,r_i\!\!\!\!\!\times,\ldots,
r_j\!\!\!\!\!\times,\ldots,r_k,r_i+r_j\rangle.
}}
Next from $(\Sigma_+ P_- - P_+ \Sigma_-)$ we get 
two terms which combine nicely into
\eqn\bhg{\eqalign{
&T_3 =- {1 \over \pi^4} \sum_{i=1}^k \sqrt{ (r_0 + r_i)^4 r_i \over r_0}
\sum_{m=-\infty}^\infty
\int_0^{r_0 + r_i} ds { \sin^2(\pi m {r_0 \over r_0 + r_i})
\sin^2(\pi m {s \over r_0 + r_i}) \over m (m - n {r_0 + r_i \over r_0})}
\cr
&\qquad\qquad\qquad\qquad\qquad\times
\left[ {1 \over m} + {1 \over m - n {r_0 + r_i \over r_0}} \right]
|r_0+r_i-s;s;r_1,\ldots,r_i\!\!\!\!\!\times,\ldots,r_k\rangle,
}}
and from $- P_+ \Sigma_-$ we get an additional term which is exactly
\eqn\aaa{
T_4 = - T_2.
}

We can simplify $T_3$ by using the sum
(B.4) with $a=0$, $b = n {r_0 + r_i \over r_0}$, but
we must consider the cases $s < r_0$ and $s > r_0$ separately, using
$y_1 = {r_0 \over r_0 + r_i}, y_2 = {s \over r_0 + r_i}$
in the former and $y_1 = 1 - {r_0 \over r_0 + r_i}$, $y_2 =
1 - {s \over r_0 + r_i}$ in the latter.  After considerable
simplification, we that
\eqn\aaa{\eqalign{
&T_3=- { \sqrt{r_0 } \over \pi^2 n}
\sum_{i=1}^k \sqrt{r_i} \int_0^{r_0 + r_i} ds
\Bigg[
{ (r_0 - s) {\textstyle{\sin^2(\pi n {s \over r_0})}}
 } \theta(r_0 - s)
\cr
&\qquad\qquad
+ { (s-r_i) {\textstyle{\sin^2(\pi n {s - r_i \over r_0})}}
} \theta(s - r_0)\Bigg] |r_0+r_i-s;s;r_1,\ldots,r_i\!\!\!\!\!\times,\ldots,
r_k\rangle.
}}
After shifting the variable of integration in the second term
it becomes manifest that $T_3 = - T_1$.
Adding up all of these terms therefore gives $T_1 + T_2 + T_3 + T_4 = 0$!

It remains to collect the large number of terms of the first
type which arise.  After several pages of algebra, these can be
shown to all cancel using all of the various ingredients used
in the previous stages of the proof.
We do not show all of the steps since the formulas
are exceedingly long and no new tricks are required.

\appendix{B}{Sums}

Throughout this appendix, $a,b,c$ denote non-integer (though
otherwise arbitrary complex) numbers, and $y \in [0,1]$.
We start with the basic sum
\eqn\basicsum{\eqalign{
\sum_{m=-\infty}^\infty { e^{2 \pi i m y} \over (m-a) (m-b) (m-c)}
&= f(a,b,c) + f(b,c,a) + f(c,a,b),
\cr
f(a,b,c) &= - {\pi \csc(\pi a) \over (a-b)(a-c)} e^{\pi i a (2 y-1)}.
}}
which is easily evaluated by standard contour techniques (see for example
\KlebanovMP).
It follows that
\eqn\aaa{\eqalign{
\sum_{m=-\infty}^\infty { \sin^2(\pi m y)
 \over
(m-a)(m-b)(m-c)} &= g(a,b,c) + g(b,c,a) + g(c,a,b),\cr
g(a,b,c) &= 
{\pi \csc(\pi a) \sin(\pi a y)
\sin(\pi a (1-y)) \over (a-b)(a-c)}.
}}
By taking appropriate limits of this result one can obtain
a formula for
the sum
\eqn\aaa{
F(y;a,b) \equiv
\sum_{m=-\infty}^\infty {\sin^2(\pi m y) \over (m-a)(m-b)}
\left[ {1 \over m-a} + {1 \over m-b}\right],
}
which appears throughout appendix A.
The expression for $F(y;a,b)$ is rather lengthy so we will not write it
down, but a particularly useful special case is
\eqn\aaa{
F(y; a = n/y, b)=
{\pi^2 \over (a-b)} 
{\sin(\pi b y) \over \sin(\pi b)}
\left[2 y \cos (\pi b(1-y)) - {\sin(\pi b y) \over \sin(\pi b)}\right],
}
valid when $n$ is an integer.

Next, for $1 \ge y_1 + y_2 \ge y_1 \ge y_2 \ge 0$ we obtain from \basicsum\ 
the more complicated result
\eqn\aaa{\eqalign{
\sum_{m=-\infty}^\infty { \sin^2(\pi m y_1) \sin^2(\pi m y_2)
\over (m-a)(m-b)(m-c)}
&= h(a,b,c) + h(b,c,a) + h(c,a,b),
\cr
h(a,b,c) &= {\pi  \over 4 (a-b) (a-c)}
{\sin (\pi a y_2) \over \sin(\pi a)}
\Big[
2 \sin(\pi a (1 - y_2))\cr
&+
\sin(\pi a (1-2 y_1 + y_2)) + \sin (\pi a ( 2 y_1 + y_2 - 1))
\Big].
}}
Finally,
taking the appropriate limits of the preceding equation gives
the remarkable identity
\eqn\aaa{\eqalign{
&\sum_{m=-\infty}^\infty {\sin^2(\pi m y_1) \sin^2(\pi m y_2)
\over m (m-a)} \left[ {1 \over m} + {1 \over m-a} \right]
= {\pi^2 \over a} {\sin^2(\pi a y_1) \sin^2(\pi a y_2) \over \sin^2(\pi a)}
\cr
&\qquad+
{\pi^2 \over 2 a} {\sin(\pi a y_2) \over \sin(\pi a)}\Bigg[
(y_1 + y_2) \cos(\pi a (1 - 2 y_1 - y_2))
\cr
&\qquad\qquad\qquad\qquad\qquad
- (y_1 - y_2) \cos(\pi a (1 - 2  y_1 + y_2))
-  2 y_2 \cos(\pi a (y_2 - 1))
\Bigg].
}}

\appendix{C}{More on $\Sigma$}

Here we present the joining operator, which is the adjoint
of \spdef:
\eqn\aaa{\eqalign{
&\Sigma_- | n, r_0; r_1,\ldots,r_k\rangle
\cr
&\qquad
= \sum_{i=1}^k \sum_{m=-\infty}^\infty
\sqrt{(r_0 + r_i)^3 r_i \over r_0}
{\sin^2(\pi m {r_0 \over r_0 + r_i}) \over
\pi^2 (m- n {r_0 + r_i \over r_0})^2}
|m,r_0 + r_i;r_1,\ldots,r_i\!\!\!\!\!\times,\ldots,r_k\rangle
\cr
&\qquad + \ha \sum_{i \ne j}
\sqrt{r_i r_j(r_i + r_j)}
|n,r_0; r_1,\ldots,r_i\!\!\!\!\!\times,\ldots,r_j\!\!\!\!\!\times,
\ldots,r_k,r_i+r_j\rangle,\cr
&\Sigma_- |s_1;s_2;r_1,\ldots,r_k\rangle =\cr
&\qquad 
\sum_{m=-\infty}^\infty {(-1)^m (s_1 + s_2)^{3/2} \over \pi^2 m^2}
{\textstyle{\sin(\pi m {s_1 \over s_1 + s_2})}}
{\textstyle{ \sin(\pi m {s_2 \over s_1 +
s_2})}} |m,s_1+s_2;r_1,\ldots,r_k\rangle\cr
&\qquad+ \sum_{i=1}^k   \sqrt{r_i} \Big[ s_1 
|s_1 + r_i; s_2;r_1,\ldots,r_i\!\!\!\!\!\times,\ldots,r_k\rangle
+ s_2  |s_1;
s_2 + r_i; r_1,\ldots,r_i\!\!\!\!\!\times,\ldots,r_k\rangle
\Big]\cr
&\qquad + \ha \sum_{i \ne j} \sqrt{r_i r_j (r_i + r_j)}
|n,r_0; r_1,\ldots,r_i\!\!\!\!\!\times,\ldots,r_j\!\!\!\!\!\times,
\ldots,r_k,r_i+r_j\rangle.
}}

\listrefs

\end